\documentclass[letterpaper]{JHEP3}

\newcommand{\beq}{\begin{equation}}
\newcommand{\eeq}{\end{equation}}

\usepackage{graphicx}
\usepackage{amsmath}
\usepackage{amsfonts}
\usepackage{epsfig}

\title{Realization of Center Symmetry in Two Adjoint Flavor Large-N Yang-Mills}

\author{Simon Catterall \\
Department of Physics, Syracuse University, Syracuse, NY 13244, USA \\
E-mail: \email{smc@phy.syr.edu}
}
\author{Richard Galvez \\
Department of Physics, Syracuse University, Syracuse NY 13244, USA\\
E-mail: \email{ragalvez@syr.edu}
}
\author{Mithat \"{U}nsal\\
SLAC and Physics Department, Stanford University, Stanford, CA 94305, USA\\
E-mail: \email{unsal@slac.stanford.edu}
}
\date{}                                        
\preprint{SLAC-PUB-14161}

\abstract{
We report on the results of numerical simulations of
$SU(N)$ lattice Yang Mills with two flavors of 
(light) Wilson fermion in the adjoint representation. We analytically and numerically address the question of center symmetry realization  on lattices with $\Gamma$ sites in each direction 
in the large-$N$ limit.  We show, by a weak coupling calculation that, for massless fermions, 
center symmetry realization  is independent of $\Gamma$, and is unbroken.  Then, we extend 
our result   by conducting simulations at non zero mass and finite
gauge coupling.  
Our results indicate that center symmetry is intact for a range of
fermion mass in the vicinity of the
critical line on lattices of volume $2^4$. This
observation makes it possible to compute infinite volume
physical observables using small volume simulations 
in the limit $N\to\infty$, with possible applications to the
determination of the conformal window in gauge theories with adjoint fermions.  
}

\begin{document}

\section{Introduction}
Large-$N$  gauge theories compactified on a torus have 
properties independent  
of the compactification radii 
provided {\it i)} center symmetry and {\it ii)} translation symmetry 
are not spontaneously broken \cite{Eguchi:1982nm, Yaffe:1981vf,  Bhanot:1982sh, GonzalezArroyo:1982hz,Kovtun:2007py}. 
We refer to this property  as large-$N$ volume independence. 
In a lattice regularized theory, the reduction to a one-site model is known as ``Eguchi-Kawai (EK) or large-$N$  reduction".
Volume independence is a property of  both confining   and 
conformal     field theories provided  the necessary symmetries are 
satisfied \cite{UY10, Poppitz:2010bt}.

If valid, large-$N$ reduction may have practical benefits in lattice gauge theory. For example, the 
computational cost of simulating a $L^4$ lattice (excluding the effects of
critical slowing down) grows as
$L^{5}$ while the computational cost of simulating at 
large-$N$ naively grows only as $N^{\frac{7}{2}}$. 
These observations are
particularly important in theories with large finite
volume effects such as conformal or near conformal theories. 
One such example of current interest is the {\it minimal walking theory} (MWT)
which has been proposed as a 
technicolor model allowing for a dynamical breaking
of electroweak symmetry 
\cite{Sannino:2004qp,Dietrich:2006cm,Evans:2005pu,Dietrich:2005jn,Foadi:2007ue,Antola:2010nt}. This model employs two Dirac  (four Weyl)  flavors of
adjoint fermion in an $SU(2)$ gauge theory and is thought
to be conformal or near conformal in the infrared. A near conformal
behavior is believed to be a necessary ingredient for constructing a
realistic ``walking" gauge theory capable of breaking the
electroweak symmetry of the Standard Model while respecting the
bounds on new physics implied by electroweak precision measurements
at LEP. Because of these features,
this model has been studied extensively in
recent years by the lattice community \cite{Catterall:2007yx,Catterall:2008qk,DelDebbio:2010hx,DelDebbio:2010hu,DelDebbio:2009fd,Bursa:2009we,Bursa:2009tj,Hietanen:2009zz,Hietanen:2009az,
Catterall:2009sb}.

For four dimensional confining gauge theories,  
it was usually believed that  large-$N$ 
reduction  should be valid only above a critical size, $L>L_{\rm c}$, below which 
center symmetry would break spontaneously invalidating the equivalence. For example, 
for pure YM theory, $L_{\rm c} \sim \Lambda^{-1} $ is about  a few fermi \cite{Kiskis:2003rd}.  
This understanding changed in recent years \cite{Kovtun:2007py,Unsal:2008ch,Bringoltz:2009kb}.   Ref.\cite{Kovtun:2007py} has shown  
that Yang-Mills theory with multiple adjoint fermions [QCD(adj)],  
where the fermions are endowed with (non-thermal) periodic boundary conditions,
satisfies volume independence down to arbitrarily small volumes.
Another  way to have  a working volume independence 
is to  use double-trace deformations, 
by explicitly adding the modulus square of Wilson line operators \cite{Unsal:2008ch}. This suppresses the unwanted breaking, without altering the leading large-$N$ dynamics of the theory. In this sense, the deformed theory also   overcomes the $L_{\rm c}$  ``impasse" of pure theory. 
\footnote{Earlier modifications   known as quenched-- \cite{ Bhanot:1982sh}
    and twisted-- EK models     \cite{GonzalezArroyo:1982hz}
    have recently been shown to fail due to nonperturbative effects
    \cite{Bietenholz:2006cz,Teper:2006sp, Azeyanagi:2007su, Bringoltz:2008av}.}

Switching back to  the minimal walking theory, one of the 
most pressing questions to resolve using non-perturbative studies
is to clarify whether this model does indeed lie within the
so-called conformal window or whether it lies just outside this window
and perhaps can serve  as an example of a walking gauge theory. 
To distinguish conformal from near conformal behavior it would
seem that very large volume simulations would be
necessary. 
Furthermore, since the conformal window for models with
adjoint fermions  is expected,   at leading order in $N$, 
to be independent of 
$N$,  the question can  equally be phrased  in a general
$N_{\rm f}$-flavor $SU(N)$ gauge theory. 
Continuum analysis indicates that this 
class of theories, endowed with periodic boundary conditions,   obey volume independence, regardless of their long-distance behavior. 
Thus, the question of whether $N_{\rm f}=2$ theory 
is conformal or confining can be addressed on  small  $2^4$ or even  $1^4$ lattices \cite{UY10}. 
Since volume reduction depends on the property of center
symmetry it is important to see, nonperturbatively,  whether  the large-$N$ theory
is center symmetric  in  the limit of small  volumes. This examination  will
be the main goal of this work.

As mentioned above, it is well known that center symmetry 
spontaneously breaks in large-$N$
pure Yang Mills theory for dimensions greater than two \cite{Bhanot:1982sh, Kiskis:2003rd}. 
However,  a one-loop continuum analysis   
on $R^3 \times S^1$ (which generalizes to 
arbitrary toroidal compactifications)  has shown that 
the center symmetry will be restored if the theory is coupled to one or more flavors of
(light or massless) adjoint\footnote{With fundamental fermions, center symmetry is  explicitly broken, but it  still is an approximate symmetry in the  $N_{\rm f} $=fixed as $N \rightarrow \infty$ limit. Volume independence is valid in the confined phase of such a theory. This domain can be extended 
to arbitrarily small radii   by the addition of double-trace deformations\cite{Unsal:2008ch}. }
Weyl fermion \cite{Kovtun:2007py}. 
See, also  Refs.\cite{Cossu:2009sq, Bedaque:2009md,  Bringoltz:2009mi, Bringoltz:2009fj, Poppitz:2009fm, Myers:2008zm, Meisinger:2009ne} for related work. 

Ref.\cite{Kovtun:2007py} also conjectured that 
a single-site version of QCD(Adj), which is just the EK matrix model  augmented with adjoint representation Grassmann variables, 
will reproduce the leading large-$N$ behavior of all expectation values, and zero-momentum connected correlators of single-trace observables in infinite volume QCD(Adj).   This conjecture received strong numerical confirmation 
in single flavor simulations in \cite{ Bringoltz:2009kb} for Wilson fermions,  in one-loop lattice perturbative analysis as in  
\cite{Hietanen:2009ex} using overlap fermions, and 
in a one-loop analysis and simulations using the one-site theory  with Wilson fermions in  
\cite{Azeyanagi:2010ne}.   
 
Our goal in this work is to extend   the one flavor result of 
\cite{Bringoltz:2009kb}, to 
two flavors  in preparation for studies of
the conformal window in the MWT theory.

\section{Lattice action and methods}

We employed a $2^4$ lattice volume in our work.  The reason we do so is to take advantage of both volume and $N$ scaling simultaneously, as we are working with relatively smaller $N$ than that used in \cite{Bringoltz:2009kb, Azeyanagi:2010ne}. 
The lattice action we employ consists of the usual Wilson plaquette
term 
\begin{equation}
S_G=-\frac{\beta}{2}\sum_x\sum_{\mu>\nu}{\rm Re}{\rm Tr}\left(
U_\mu(x)U_\nu(x+\mu)U^\dagger_\mu(x+\nu)U^\dagger_\nu(x)\right) \end{equation}
together with the Wilson action for two Dirac quarks in the adjoint
representation
\begin{eqnarray}
S_F&=&-\frac{1}{2}
\sum_x\sum_\mu\overline{\psi}(x)\left(
V_\mu(x)\left(I-\gamma_\mu\right)\psi(x+\mu)+
V^\dagger_\mu(x-\mu)\left(I+\gamma_\mu\right)\psi(x-\mu)
\right)\\
&+&\sum_x  \left(m+4\right) \overline{\psi}(x)\psi(x) \ ,
\end{eqnarray}
where the symmetric links are given by
\begin{equation}
V^{ab}_\mu(x)={\rm Tr}\left(S^aU_\mu(x)S^bU^T_\mu(x)\right) \ ,
\end{equation}
and the matrices $S^a,a=1,2,3$ are the usual Pauli matrices.
We use the usual HMC algorithm \cite{hmc} to simulate this model at a variety
of 't Hooft couplings $\lambda=g^2N$ and bare quark masses $m$. Periodic
boundary conditions are used for all lattice directions.

\section{Weak coupling analysis of center symmetry  on  $\Gamma^4$ lattice} 
Before showing the results of our numerical simulations,  it is
instructive to consider the question of
center symmetry realization in a weak coupling 
lattice perturbation theory analysis.  We consider a four-dimensional lattice  with $\Gamma^4$ sites, labeled as  $L^{\Gamma}$. Generalization to asymmetric lattices
with $\Gamma_{\mu}$ sites in the $\mu$-th directions is obvious.
Let us label the  Wilson line  (none of the circles is thermal, with a slight abuse of language, we use  the Wilson  and Polyakov line interchangeably)  along the $\mu$-th direction 
as $P_{\mu}$, given by
\begin{equation}
P_{\mu}(x) = U_{\mu}(x) U_{\mu}(x+ e_{\mu}) U_{\mu}(x+ 2e_{\mu}) \ldots  U_{\mu}(x+ 
(\Gamma-1) e_{\mu})
\label{Pline}
\end{equation}
$P_{\mu}$ is gauge covariant and its trace is gauge invariant. 
      The space of classical vacua is parameterized as  a space of commuting (diagonal) Wilson lines,
    $[P_{\mu}, P_{\nu}]=0$: 
   \begin{equation}
P_{\mu} = {\rm Diag}  \left( e^{i \theta_{\mu}^1}, e^{i \theta_{\mu}^2},  \ldots, 
e^{i \theta_{\mu}^N}  \right)
\end{equation}
In the weak coupling regime, it is 
easy to evaluate the one-loop potential by integrating out the heavy modes. 
Let us first  find the spectrum of gauge and fermionic fluctuations in the background of commuting Wilson lines.  Our calculation generalizes section 2 of Ref.\cite{Poppitz:2009fm}.
%and  one-site calculation of \cite{Azeyanagi:2010ne}. 

The spectrum of gauge fluctuations in the background of commuting Wilson lines is  
\begin{eqnarray}
&& M^2_g(k_\mu, \theta_{\mu}^{ij} )= \sum_{\mu=1}^{4}  \left[\frac{2}{a} \sin \left(\frac{2 \pi k_{\mu}  +  \theta_{\mu}^{ij} }{2\Gamma} \right) \right]^2, \qquad  
k_{\mu}=1, \ldots \Gamma, \; \;  i, j=1, \ldots N~
\label{spectrum}
\end{eqnarray}
where $\theta_{\mu}^{ij} \equiv  \theta_{\mu}^{i} - \theta_{\mu}^{j}$.
And analogously, for Wilson fermions with bare mass $m$ and Wilson parameter $r=1$, 
the spectrum of fermionic fluctuations is given by 
\begin{eqnarray}
M_f^2[k_{\mu}, \theta_{\mu}^{ij}, m]&&= \frac{1}{a^2} \sum_{\mu=1}^{4} \sin^2\left(\frac{2 \pi k_{\mu}  +  \theta_{\mu}^{ij} }{\Gamma} \right)+ 
 \left[ m  + \frac{2}{a} \sum_{\mu=1}^{4}  \sin^2 \left(\frac{2 \pi k_{\mu}  +  \theta_{\mu}^{ij} }{2\Gamma} \right)  \right]^2 
\end{eqnarray}
Both of these formulas are quite intuitive:
Setting $\theta_{\mu}^{ij} = 0$ gives the usual spectrum of gauge bosons
and Wilson-fermions (with $r=1$) in   lattice gauge theory. 
Setting  $\Gamma= 1$, the bosonic formula  gives the distance between the eigenvalues of the Wilson line. The fermionic one is similar, but  differs due to the fermion dispersion relation on the lattice. 
Note that lattice momenta $k_\mu$ and eigenvalue difference  
$\theta_{\mu}^{ij}$ appear on the same footing. 
 
The one-loop action induced by  bosonic and fermionic  
fluctuations  on $L^\Gamma$  yields
\begin{eqnarray}
S_{\rm 1-loop}[\Gamma, \theta_{\mu}^{ij}, m] &&=  2  \sum_{i<j}   \sum_{\vec k \in    L^{\Gamma} }   
\;  \log{\left (M^2_g(k_\mu, \theta_{\mu}^{ij} )\right)}  -  4N_f      \sum_{i<j}   \sum_{\vec k \in    L^{\Gamma} }  \; 
\log{\left(M_f^2(k_{\mu}, \theta_{\mu}^{ij}, m)\right)}
 \label{sin2} \qquad 
\end{eqnarray}
This formula captures all the interesting limits, including the 1-site lattice theory (EK-version of QCD(adj)) and the 
$\Gamma=\infty$ infinite lattice limit, as well as the continuum limit.   
Below, we concentrate on the case where the
bare mass is set to zero, $m=0$. 

For $N_f=0$ $\Gamma=1$ we recover the result  of  Ref. 
\cite{Bhanot:1982sh} in which a spontaneous breaking of center symmetry occurs. If we take 
the continuum limit by using the usual scaling, 
\begin{equation}
\Gamma \rightarrow \infty, \qquad a \rightarrow 0,  \qquad L= \Gamma a= {\rm fixed}~,
\label{cont}
  \end{equation}
 with  $L \Lambda_{\rm YM } \ll1$ ($\Lambda_{\rm YM} $ is the strong-coupling scale of the theory),  one must   reproduce the  continuum  one-loop result  from perturbation theory.  Indeed, using  
 \begin{equation}
 \lim_{a \rightarrow 0} \frac{2}{a} \sin \left(\frac{2 \pi k_{\mu}  +  \theta_{\mu}^{ij} }{2\Gamma}\right) =  \left( 2 \pi k_{\mu}  +  \theta_{\mu}^{ij}  \right)L^{-1} \; ,
 \label{conti}
 \end{equation} 
the one-loop potential in pure YM theory on continuum small $T^4$ is
produced.   This expression can equivalently be written by  using Poisson resummation in terms of gauge invariant Wilson lines and the result is 
 %  \begin{eqnarray}
 $ S_{\rm 1-loop}^{\rm YM}[P_{\mu} ]=    
  -  \frac{1}{\pi^2}
    \sum_{ \vec n \in {\mathbb Z}^4 \setminus \{{\bf  0}\}} \frac{1}{ |\vec n|^4} 
      | {\rm Tr} ( P_{1}^{n_1}   \ldots  P_{4}^{n_4}) |^2    $.
% \label{MMP2} \end{eqnarray}
Thus for Yang-Mills theory on small continuum $T^4$, 
since the masses of Wilson lines are all negative, 
the  center symmetry  is broken.

The case of $\Gamma=1$ and $N_f \geq 0.5$ is the 
full EK reduced version of QCD(adj) simulated in 
\cite{Bringoltz:2009kb, Azeyanagi:2010ne}.  As opposed to gauge fluctuations which generate eigenvalue attraction in any domain where one-loop analysis is reliable, the adjoint fermions induce eigenvalue repulsion. This is the  adjoint fermion-induced  center-stabilization mechanism of Ref.\cite{Kovtun:2007py}.  
If we take  the (naive) continuum limit by using the  scaling (\ref{cont}),  then the result  
reduces to 
\begin{equation}S_{\rm 1-loop}^{\rm QCD}[P_{\mu} ]  = (1-2N_f) S_{\rm 1-loop}^{\rm YM}[P_{\mu} ] =    \frac{1}{\pi^2}
    \sum_{ \vec n \in {\mathbb Z}^4 \setminus \{{\bf  0}\}} \frac{  (-1+2N_f) }{ |\vec n|^4} 
      | {\rm Tr} ( P_{1}^{n_1}   \ldots  P_{4}^{n_4}) |^2
      \label{potc}
\end{equation}
in agreement with  Ref.\cite{Kovtun:2007py}. 
This implies, in the massless continuum limit,  
that all Wilson lines have positive masses and the 
center symmetry is unbroken.

For arbitrary $\Gamma$, the  Poisson resummation (which can be done numerically)   leads to an  analog of the  one-loop action  \ref{potc}. This yields,  for massless fermions, that the  center symmetry realization is independent of $\Gamma$, and is unbroken.  

The above analysis is at weak coupling in lattice perturbation theory.  In the next section, we would like to check that the conclusion regarding unbroken center symmetry can be generalized to finite couplings  where a perturbative analysis is inapplicable, and non-zero fermion mass.

\section{Numerical results}
To look for a breaking of center symmetry we monitored a sequence of
order parameters corresponding to correlators of Polyakov lines in
different directions. The Polyakov line is defined in the usual
way as in  \ref{Pline}, \footnote{We average this quantity over all 8 points in the volume
orthogonal to the line direction $\mu$.}
%\begin{equation} P_\mu(x)=\prod_{t=0}^{L-1} U_\mu(x+t\hat{\mu})\end{equation}
while the line correlators we measure are given by  
\begin{eqnarray}
M^{(1)}_{\mu\nu}&=&P_\mu(x)P_\nu(x)\\
M^{(2)}_{\mu\nu}&=&P_\mu(x)P_\nu^\dagger(x)
\end{eqnarray}

\begin{figure}[] 
\begin{centering}  
\includegraphics[width=0.75\textwidth]{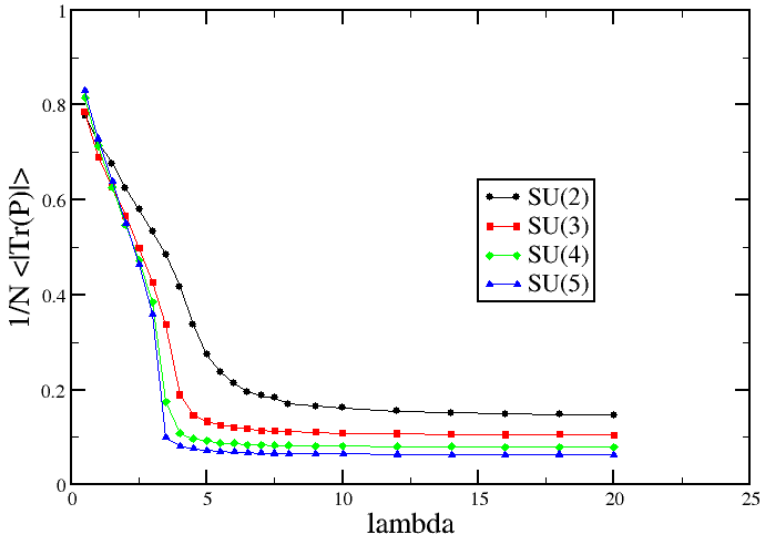} 
\caption{Simple Polyakov line in the quenched theory vs 't Hooft coupling
$\lambda$}
\label{quenchedpoly} 
\end{centering} 
\end{figure}
\begin{figure}[] 
\begin{centering}  
\includegraphics[width=0.75\textwidth]{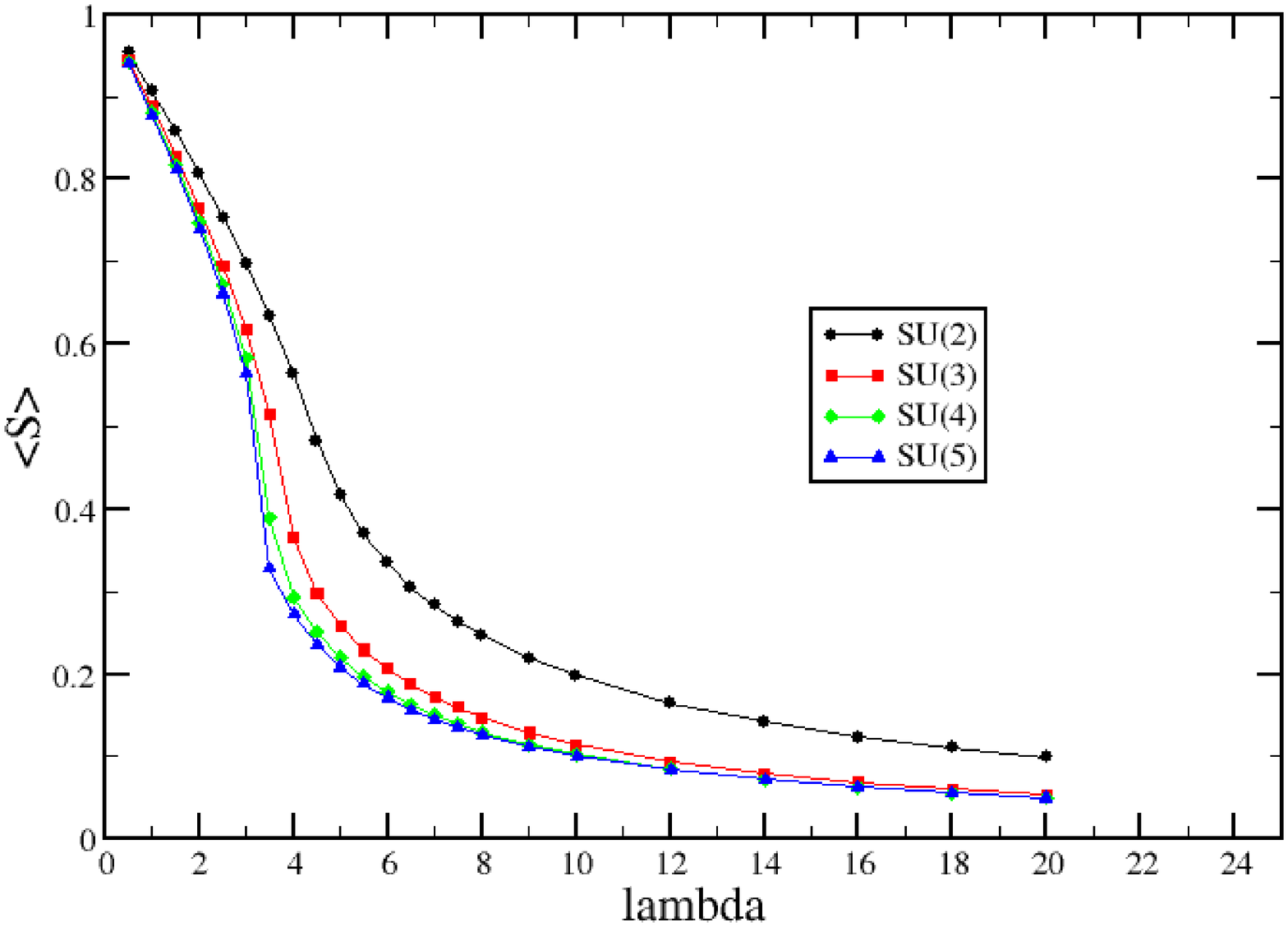} 
\caption{Plaquette action in the quenched theory vs 't Hooft coupling}
\label{quenchedaction} 
\end{centering} 
\end{figure}

It is interesting to consider first the quenched  model (with infinitely heavy fermions) 
\footnote{Here, and throughout the text,  we use the word
quenched to denote the theory without dynamical fermions. This is
different from 
Quenched-EK \cite{Bhanot:1982sh} which refers to the freezing of the eigenvalues at the roots of unity. }
 in order to
contrast the behavior of the theory
coupled to dynamical adjoint fermions. Figure~\ref{quenchedpoly}. shows
a plot of the ensemble average
of the absolute value of $\frac{1}{N}{\rm Tr}(P_1)$.
The data is plotted as a function of the
't Hooft coupling $\lambda$ for values of $N=2-5$. Clearly for
$\lambda>\lambda_c\sim 3.0$ the curves fall towards
the x-axis with larger $N$ consistent with a vanishing of
the expectation value of the line $<\frac{1}{N}{\rm Tr}(P_\mu)>=0$ in the large $N$ limit.
This implies that center symmetry remains unbroken at large
$N$. However, a different behavior is seen for small $\lambda<\lambda_c$.
In this region the curves for different $N$ all approach a fixed
$N$-independent function $f(\lambda)$ for large
$N$. Such a behavior
indicates a spontaneous breaking of center symmetry.
This two phase structure is also seen in the mean plaquette
action which is plotted in Figure~\ref{quenchedaction}.
These results reproduce what has long been known for the  pure YM theory:
the lattice theory exists in two phases; a strong coupling lattice phase with unbroken 
center symmetry and a weak coupling phase in which the symmetry is
spontaneously broken. These two phases are thought to be separated by
a phase transition - the Gross-Witten transition which is hinted at
by a potential discontinuity which appears to develop in the mean action
for large $N$ as seen in figure.~\ref{quenchedaction}. Continuum physics can be only obtained in 
the weak coupling phase and hence volume reduction is invalid  in the 
continuum limit of the theory.

\begin{figure}[] 
\begin{centering}
\includegraphics[width=0.75\textwidth]{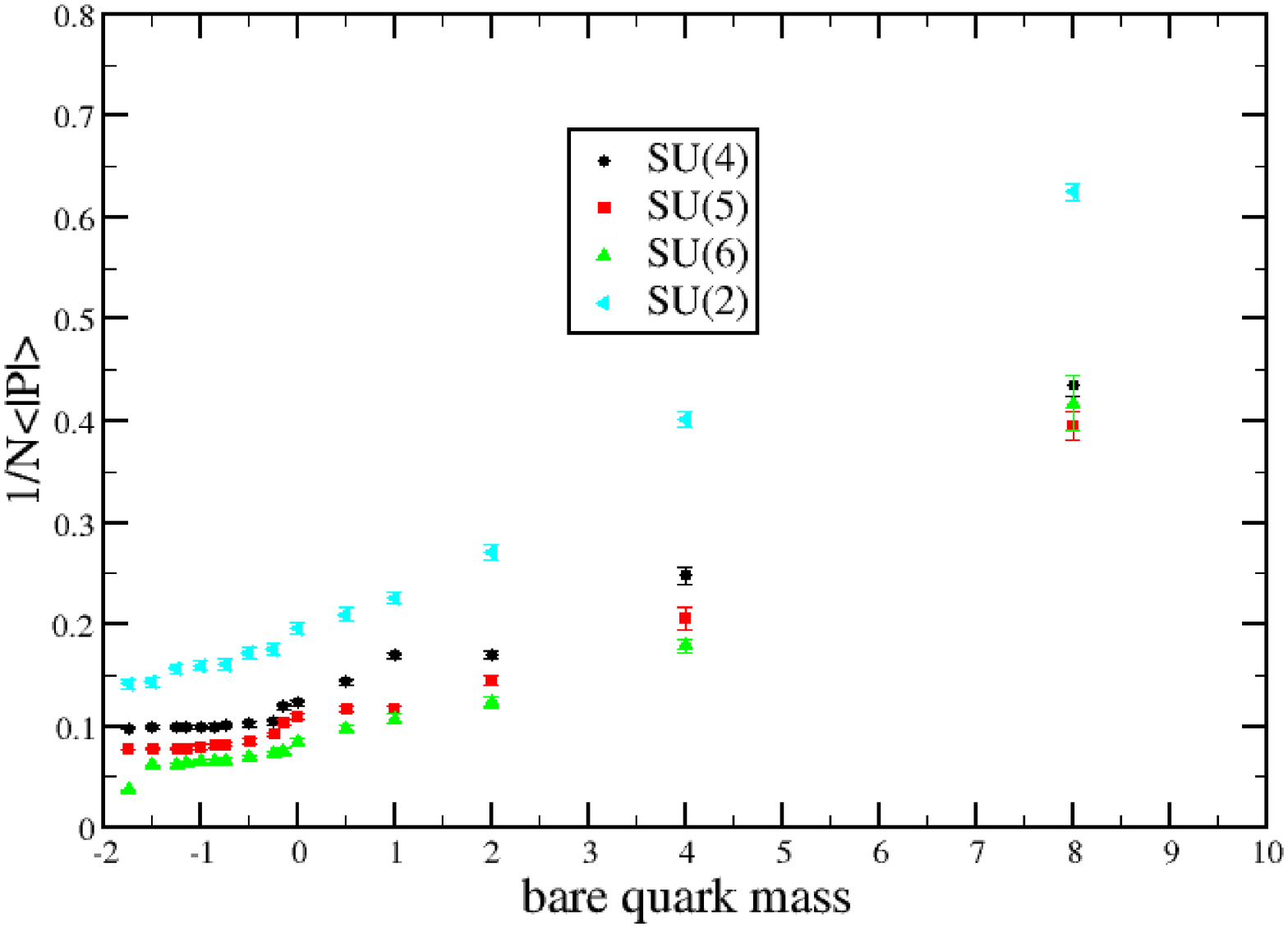}
\caption{Polyakov line vs bare quark mass for 't Hooft coupling $\lambda=0.5$} 
\label{dynpoly} 
\end{centering} 
\end{figure}

\begin{figure}[] 
\begin{centering}
\includegraphics[width=0.75\textwidth]{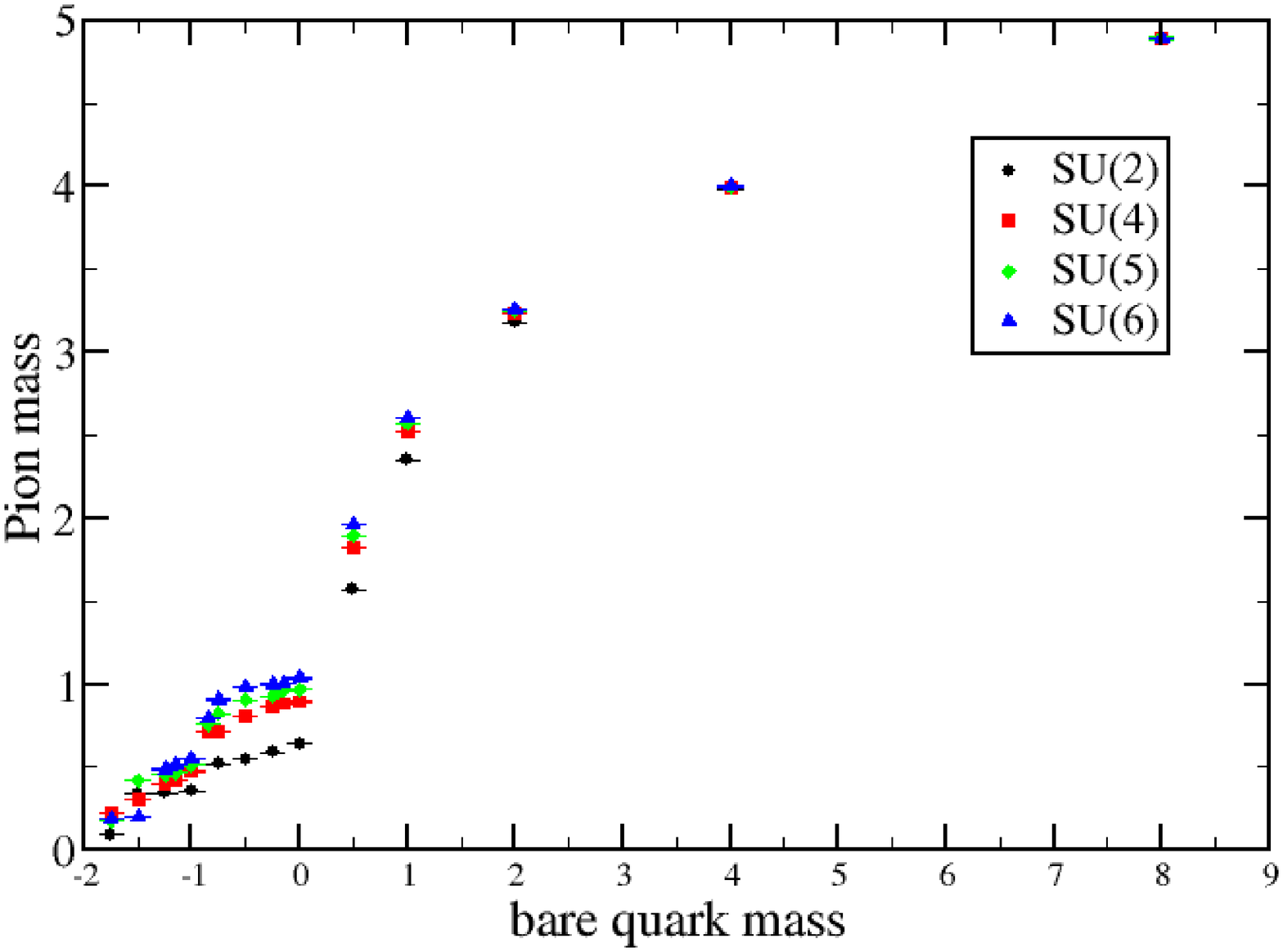}
\caption{Pion mass vs bare quark mass for 't Hooft coupling $\lambda=0.5$} 
\label{pionmass} 
\end{centering} 
\end{figure}

\begin{figure}[] 
\begin{centering}
\includegraphics[width=0.75\textwidth]{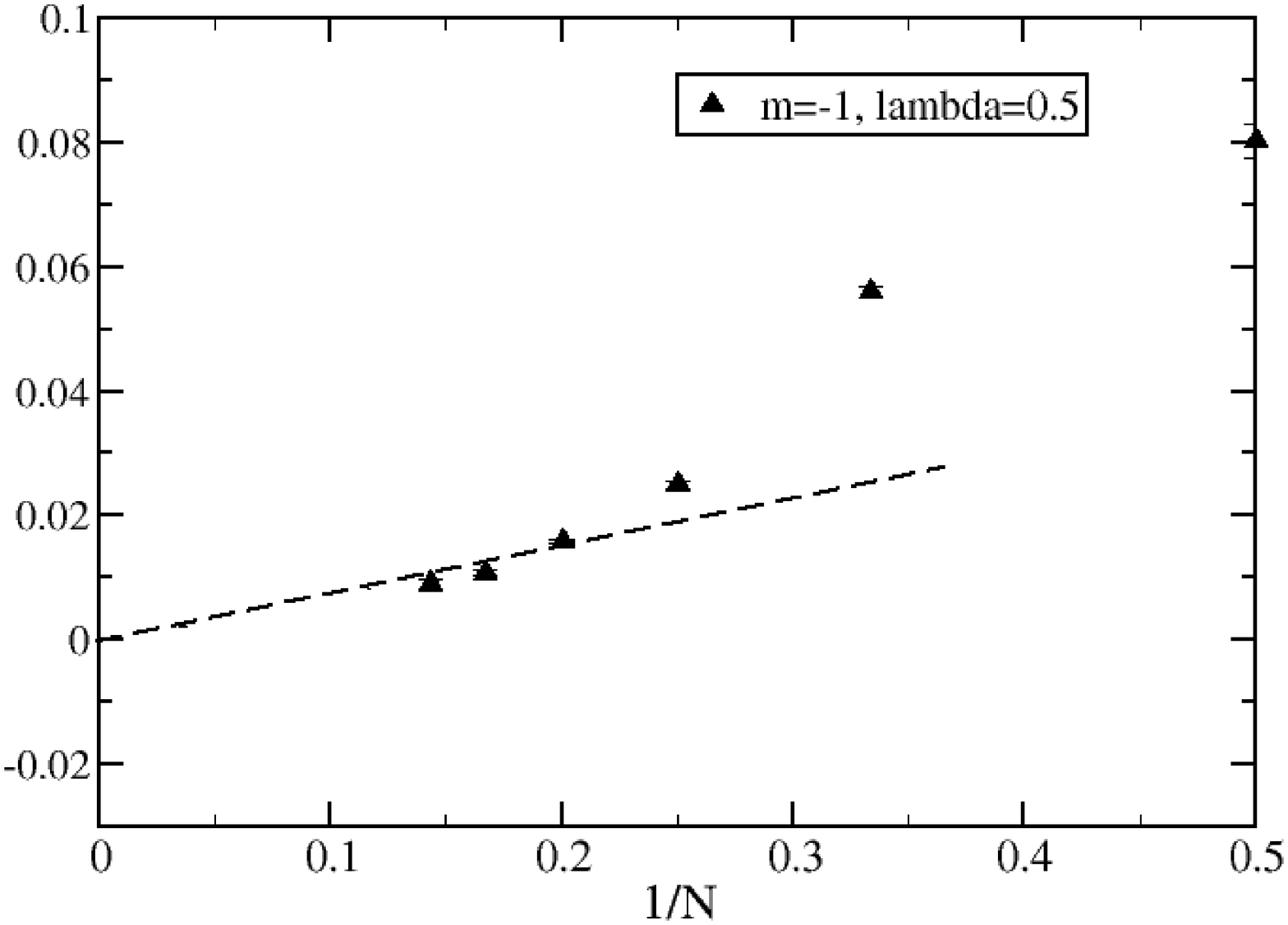}
\caption{Polyakov line vs $1/N$ for $\lambda=0.5$ and $m=-1$} 
\label{oneOverN} 
\end{centering} 
\end{figure}

\begin{figure}[] 
\begin{centering}
\includegraphics[width=0.75\textwidth]{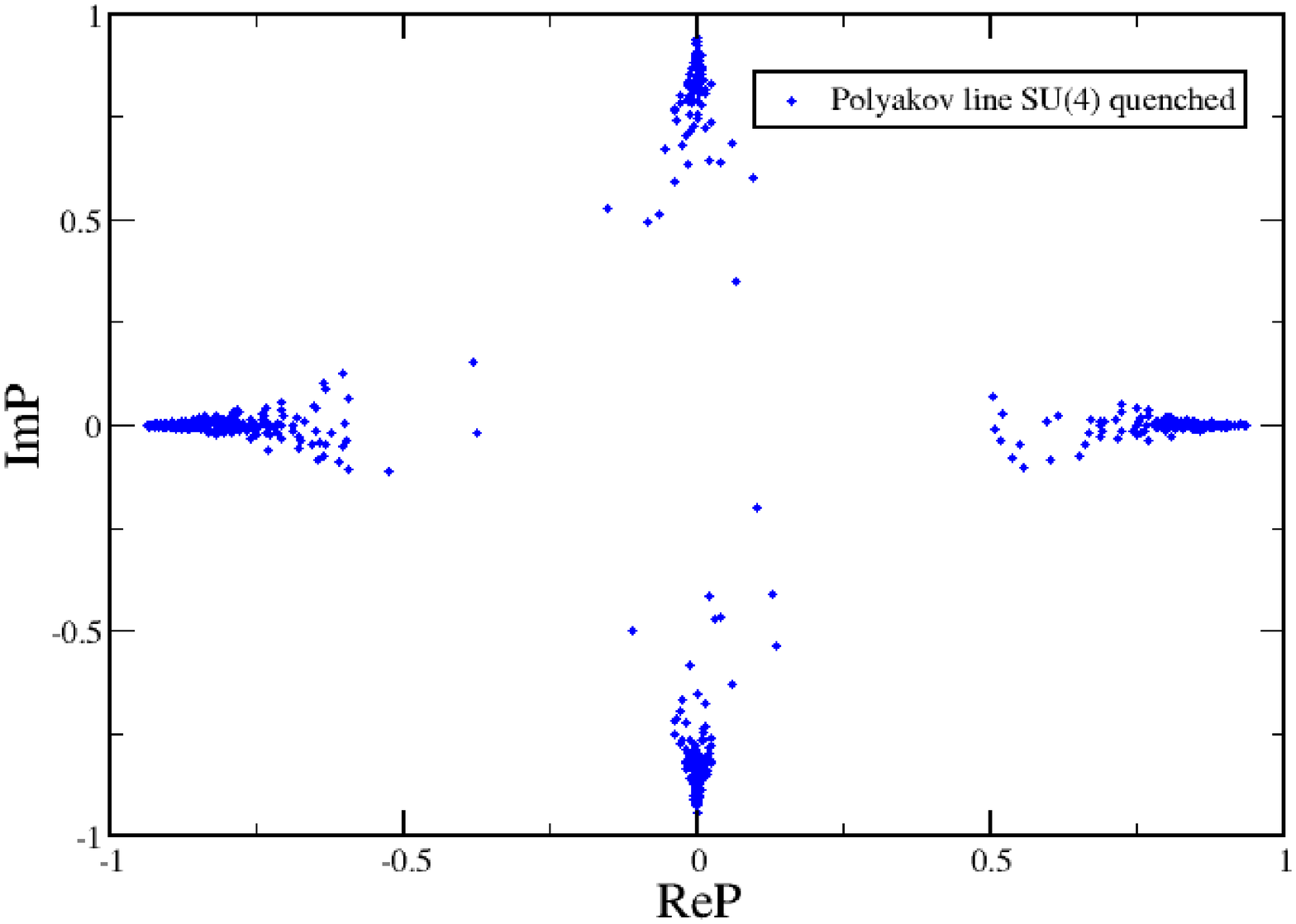}
\caption{Scatter plot of values of $P$ for quenched $SU(4)$ $\lambda=0.5$} 
\label{quenchedscatter} 
\end{centering} 
\end{figure}

\begin{figure}[] 
\begin{centering}
\includegraphics[width=0.75\textwidth]{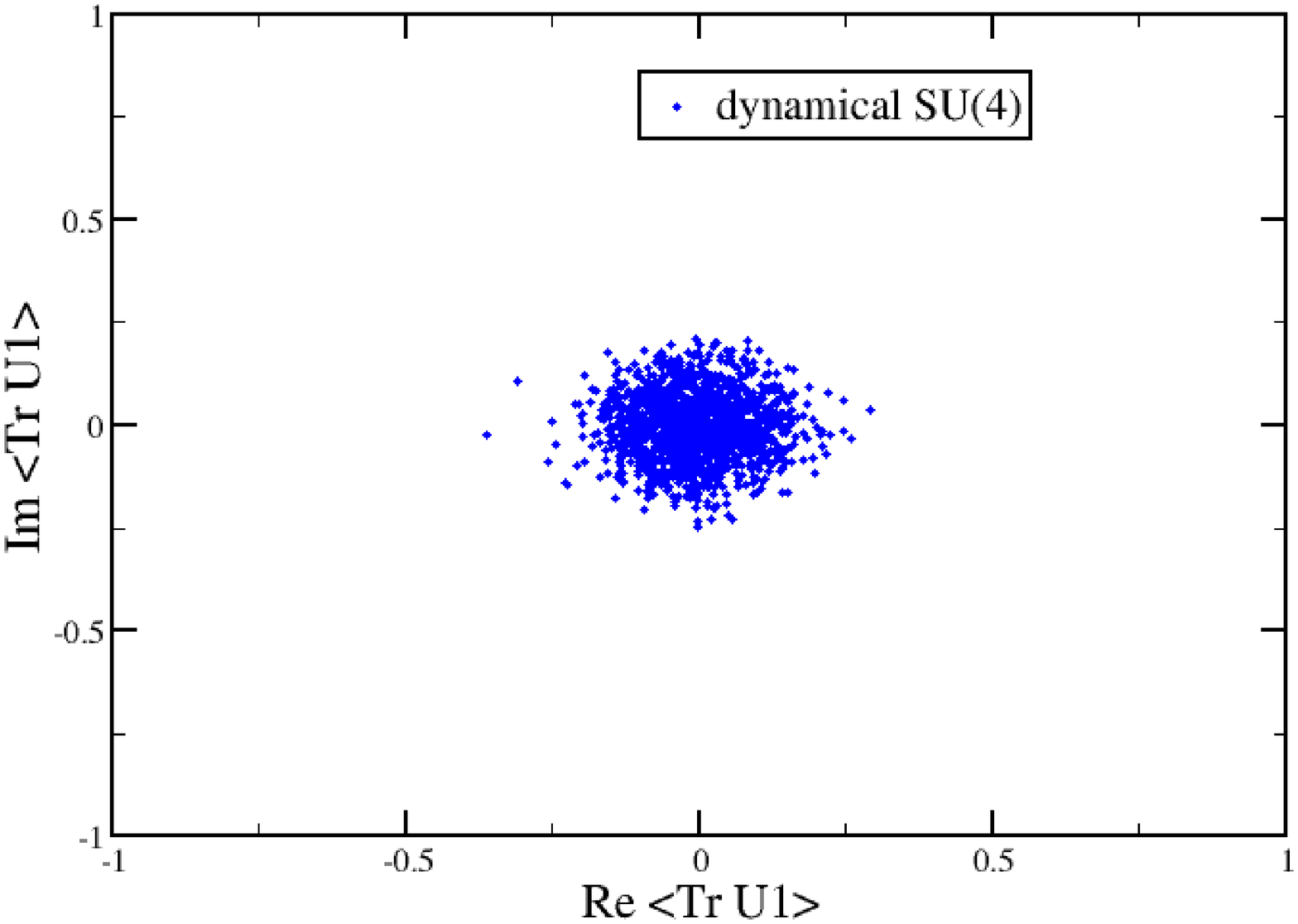}
\caption{Scatter plot of values of $P$ for dynamical $SU(4)$ $\lambda=0.5$} 
\label{dynscatter} 
\end{centering} 
\end{figure}

\begin{figure}[] 
\begin{centering}
\includegraphics[width=0.75\textwidth]{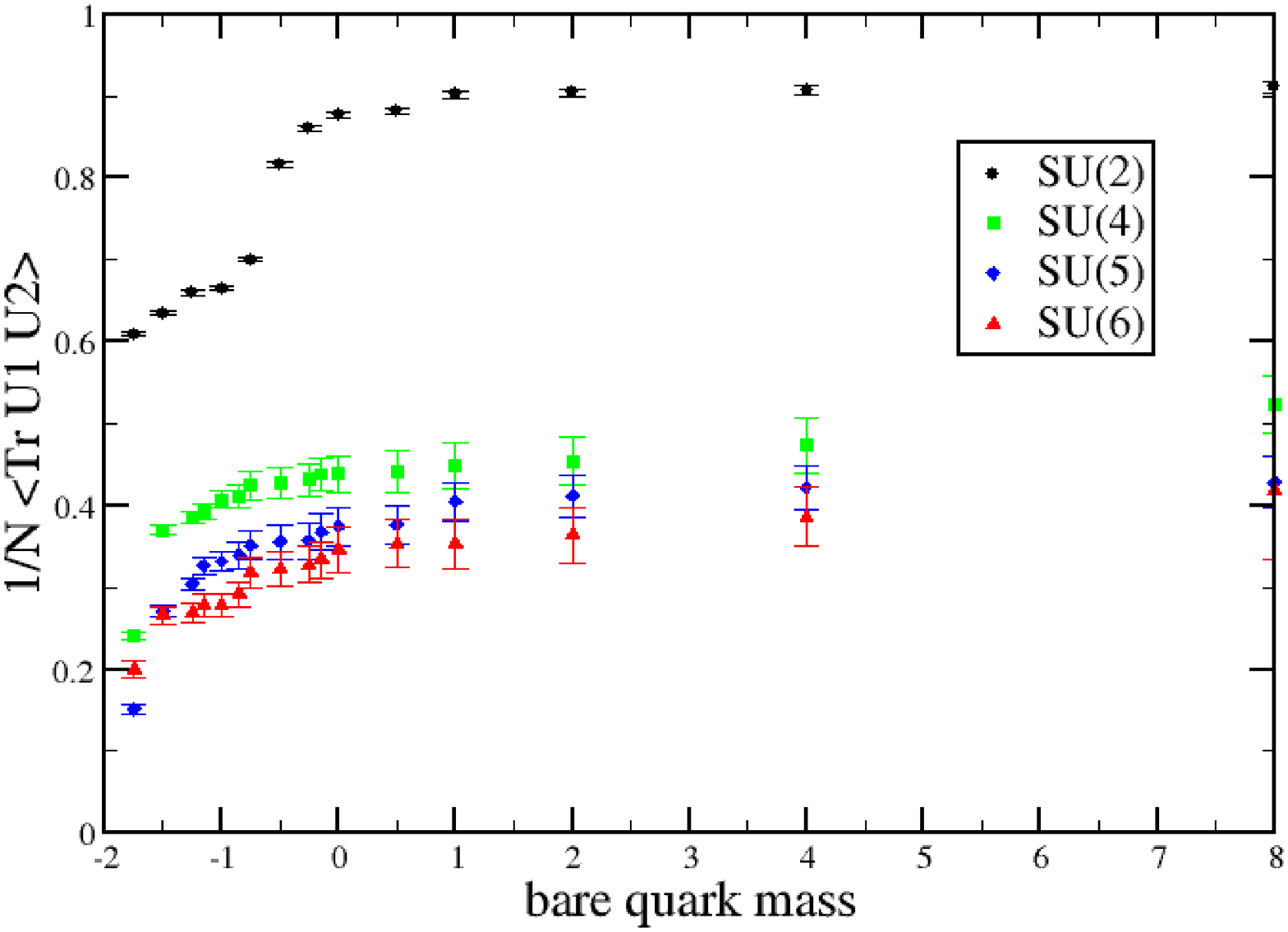}
\caption{$M^{(1)}$ vs bare quark mass for 't Hooft coupling $\lambda=0.5$} 
\label{mixpoly} 
\end{centering} 
\end{figure}

\begin{figure}[] 
\begin{centering}
\includegraphics[width=0.75\textwidth]{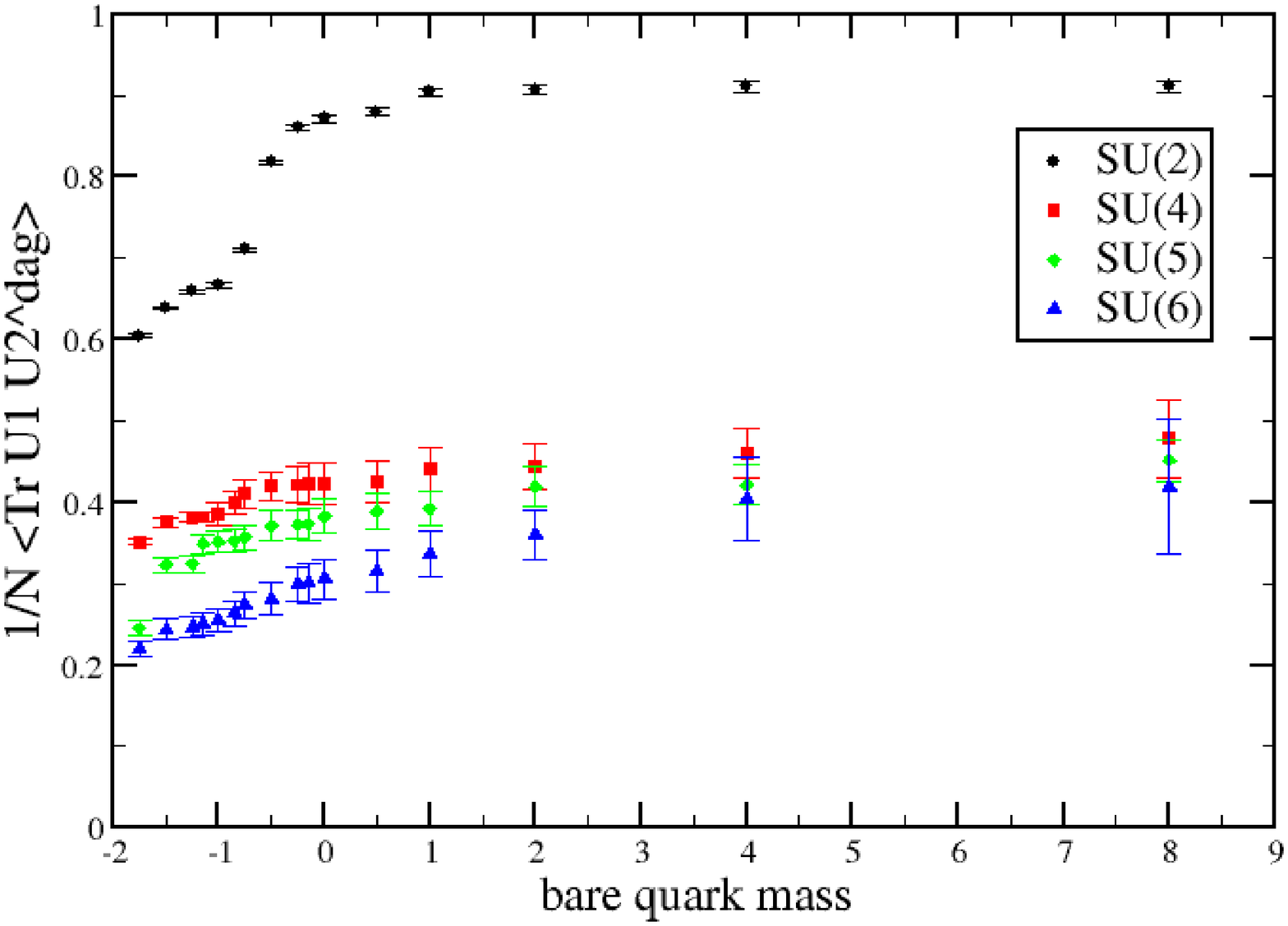}
\caption{$M^{(2)}$ vs bare quark mass for 't Hooft coupling $\lambda=0.5$} 
\label{mixpolydag} 
\end{centering} 
\end{figure}

This behavior should be contrasted with the results for the theory with
2 flavors of
dynamical adjoint fermions. Figure~\ref{dynpoly} shows a plot of the absolute
value of the Polyakov line $\frac{1}{N}|{\rm Tr}(P_1)|$
for fixed 't Hooft coupling $\lambda=0.5$ as a function of the bare 
quark mass for a range of $N=2-6$. Notice that $\lambda=0.5$
lies well within the weak coupling phase of the quenched model - a regime
in which the quenched model exhibits strong breaking of center symmetry.
The adjoint fermions seem to strongly suppress the Polyakov line and
this effect increases with larger $N$ consistent with the presence
of exact center
symmetry at large $N$.
In Figure~\ref{pionmass} we show also for this same coupling
the ``pion mass'' as
give by the logarithm of the ratio of the pion correlator at zero and
one unit of time separation $m_\pi=\log{\frac{G_\pi(0)}{G_\pi(1)}}$\footnote{
This should not be taken seriously as the real pion mass in this
theory which can only be extracted at large Euclidean times. But its value
does correlate with the position of the critical line in the model.
An equally good observable would be the number of conjugate gradient
iterations needed to invert the Dirac operator which yields a very similar
behavior with bare quark mass}. We expect the critical line to
be located close to the minimum pion mass. By comparing these two
plots it should be clear that for moderately light quark masses
the magnitude of the Polyakov line falls with increasing $N$ consistent with
a vanishing of the expectation value of $P_\mu$
in the limit $N\to\infty$. A more
quantitative measure of this is given in Figure~\ref{oneOverN} which plots
the expectation value of the Polyakov line at $\lambda=0.5$ and bare quark mass $m=-1$
as a function of $1/N$.
A linear fit to the largest $N$ points yields an intercept which is
consistent with zero and a restoration of center symmetry in the large
$N$ limit.
Further evidence in favor of center symmetry restoration comes from
comparing a scatter plot of the expectation values of the trace
of the Polyakov line
in the complex plane in both quenched and dynamical cases at $\lambda=0.5$
(in the latter case the bare fermion mass $m=-1$ placing it close to
the critical line). We show data in Figures~\ref{quenchedscatter}
and \ref{dynscatter} for the case $N=4$. The quenched data shows quite
clearly the effect of symmetry breaking -- the values of $P$ cluster
about 4 points in the complex plane corresponding to the eigenvalues
of $P$ localizing
on the fourth roots of
unity as expected in the broken phase. 
The dynamical runs show a symmetric clustering of values around
the origin as the corresponding eigenvalues spread uniformly around the
unit circle and provides strong evidence for a restoration of center symmetry.
One might worry that the data we have shown so far indeed indicates that
center symmetry is at least partially realized in the large $N$ limit but
does not exclude the possibility of more exotic breakings in which, for
example, the expectation value of the Polyakov line is vanishing in any
particular direction but that other correlators such as $M^{(1)}_{\mu\nu}$ or
$M^{(2)}_{\mu\nu}$ are non-zero - corresponding
to a locking of the values of the lines in
different directions. We find that this is not the case; Figure~\ref{mixpoly}
shows the value of $\frac{1}{N}|{\rm Tr}(M^{(1)}_{12})|$ 
versus bare quark mass for a range of
$N$ and
Figure~\ref{mixpolydag} similar curves for $\frac{1}{N}|{\rm Tr}(M^{(2)}_{12})|$. These
plots show results
which are consistent with a full realization of center symmetry 
for light quarks
in the large $N$ limit in agreement with the perturbative analysis.

\section{Conclusions}
Our analytical and
numerical results suggest that 
%strongly favor a scenario in which
the vacuum of $SU(N)$ gauge theory 
with two flavors of light adjoint Wilson fermion realizes
full center symmetry in the large $N$ limit.
This opens up the possibility of studying the (near) conformal
behavior of the $SU(2)$ minimal walking theory
by examining the large-$N$ behavior
of its $SU(N)$ generalization on small lattices.
We hope this will give a useful
additional theoretical tool for
determining whether this theory does
indeed possess an infrared conformal fixed point.

\acknowledgments 
S.M.C. is supported in part by DOE grant
DE-FG02-85ER40237. The simulations were carried out using USQCD
resources at Fermilab. 
R.G. would like to thank Ari Hietanen for useful discussions and 
 Syracuse University for the award of a STEM graduate
fellowship. M.\"U's work is supported by the
U.S.\ Department of Energy Grant DE-AC02-76SF00515. 

\bibliographystyle{JHEP}

\begin{thebibliography}{99}

%\cite{Eguchi:1982nm}
\bibitem{Eguchi:1982nm}
  T.~Eguchi and H.~Kawai,
  ``Reduction Of Dynamical Degrees Of Freedom In The Large N Gauge Theory,''
  Phys.\ Rev.\ Lett.\  {\bf 48}, 1063 (1982).
  %%CITATION = PRLTA,48,1063;%%
  
  %\cite{Yaffe:1981vf}
\bibitem{Yaffe:1981vf}
  L.~G.~Yaffe,
  ``Large N Limits As Classical Mechanics,''
  Rev.\ Mod.\ Phys.\  {\bf 54}, 407 (1982).
  %%CITATION = RMPHA,54,407;%%
  
  
%\cite{Bhanot:1982sh}
\bibitem{Bhanot:1982sh}
  G.~Bhanot, U.~M.~Heller and H.~Neuberger,
  ``The Quenched Eguchi-Kawai Model,''
  Phys.\ Lett.\  B {\bf 113}, 47 (1982).
  %%CITATION = PHLTA,B113,47;%%

%\cite{GonzalezArroyo:1982hz}
\bibitem{GonzalezArroyo:1982hz}
  A.~Gonzalez-Arroyo and M.~Okawa,
   ``The Twisted Eguchi-Kawai Model: A Reduced Model For Large N Lattice Gauge
  Theory,''
  Phys.\ Rev.\  D {\bf 27}, 2397 (1983).
  %%CITATION = PHRVA,D27,2397;%%
  
%\cite{Kovtun:2007py}
\bibitem{Kovtun:2007py}
  P.~Kovtun, M.~\"Unsal and L.~G.~Yaffe,
  ``Volume independence in large N(c) QCD-like gauge theories,''
  JHEP {\bf 0706}, 019 (2007)
  [arXiv:hep-th/0702021].
  %%CITATION = JHEPA,0706,019;%%
  

  %\cite{Unsal:2010qh}
\bibitem{UY10}
  M.~\"Unsal and L.~G.~Yaffe,
  ``Large-N volume independence in conformal and confining gauge theories,''
  arXiv:1006.2101 [hep-th].
  %%CITATION = ARXIV:1006.2101;%%



%\cite{Poppitz:2010bt}
\bibitem{Poppitz:2010bt}
  E.~Poppitz and M.~\"Unsal,
  ``AdS/CFT and large-N volume independence,''
  arXiv:1005.3519 [hep-th].
  %%CITATION = ARXIV:1005.3519;%%
   
  

  %\cite{Sannino:2004qp}
\bibitem{Sannino:2004qp}
  F.~Sannino and K.~Tuominen,
  ``Techniorientifold,''
  Phys.\ Rev.\  D {\bf 71}, 051901 (2005)
  [arXiv:hep-ph/0405209].
  %%CITATION = PHRVA,D71,051901;%%

%\cite{Dietrich:2006cm}
\bibitem{Dietrich:2006cm}
  D.~D.~Dietrich and F.~Sannino,
  ``Walking in the SU(N),''
  Phys.\ Rev.\  D {\bf 75}, 085018 (2007)
  [arXiv:hep-ph/0611341].
  %%CITATION = PHRVA,D75,085018;%%

\bibitem{Evans:2005pu}
  N.~Evans and F.~Sannino,
  ``Minimal walking technicolour, the top mass and precision electroweak
  measurements,''
  arXiv:hep-ph/0512080.
  %%CITATION = HEP-PH/0512080;%%
 %\cite{Dietrich:2005jn}
 
\bibitem{Dietrich:2005jn}
  D.~D.~Dietrich, F.~Sannino and K.~Tuominen,
  ``Light composite Higgs from higher representations versus electroweak
  precision measurements: Predictions for LHC,''
  Phys.\ Rev.\  D {\bf 72}, 055001 (2005)
  [arXiv:hep-ph/0505059].
  %%CITATION = PHRVA,D72,055001;%%
 
\bibitem{Foadi:2007ue}
  R.~Foadi, M.~T.~Frandsen, T.~A.~Ryttov and F.~Sannino,
  ``Minimal Walking Technicolor: Set Up for Collider Physics,''
  Phys.\ Rev.\  D {\bf 76}, 055005 (2007)
  [arXiv:0706.1696 [hep-ph]].
  %%CITATION = PHRVA,D76,055005;%%
%\cite{Evans:2005pu}
  
%\cite{Antola:2010nt}
\bibitem{Antola:2010nt}
  M.~Antola, S.~Di Chiara, F.~Sannino and K.~Tuominen,
  ``Minimal Super Technicolor,''
  arXiv:1001.2040 [hep-ph].
  %%CITATION = ARXIV:1001.2040;%%
%\cite{Foadi:2007ue}

%\cite{Catterall:2007yx}
\bibitem{Catterall:2007yx}
  S.~Catterall and F.~Sannino,
  ``Minimal walking on the lattice,''
  Phys.\ Rev.\  D {\bf 76}, 034504 (2007)
  [arXiv:0705.1664 [hep-lat]].
  %%CITATION = PHRVA,D76,034504;%%

%\cite{Catterall:2008qk}
\bibitem{Catterall:2008qk}
  S.~Catterall, J.~Giedt, F.~Sannino and J.~Schneible,
  ``Phase diagram of SU(2) with 2 flavors of dynamical adjoint quarks,''
  JHEP {\bf 0811}, 009 (2008)
  [arXiv:0807.0792 [hep-lat]].
  %%CITATION = JHEPA,0811,009;%%
  
%\cite{DelDebbio:2010hx}
\bibitem{DelDebbio:2010hx}
  L.~Del Debbio, B.~Lucini, A.~Patella, C.~Pica and A.~Rago,
  ``The infrared dynamics of Minimal Walking Technicolor,''
  arXiv:1004.3206 [hep-lat].
  %%CITATION = ARXIV:1004.3206;%%

%\cite{DelDebbio:2010hu}
\bibitem{DelDebbio:2010hu}
  L.~Del Debbio, B.~Lucini, A.~Patella, C.~Pica and A.~Rago,
  ``Mesonic spectroscopy of Minimal Walking Technicolor,''
  arXiv:1004.3197 [hep-lat].
  %%CITATION = ARXIV:1004.3197;%%
%\cite{Bursa:2009we}

%\cite{DelDebbio:2009fd}
\bibitem{DelDebbio:2009fd}
  L.~Del Debbio, B.~Lucini, A.~Patella, C.~Pica and A.~Rago,
  ``Conformal vs confining scenario in SU(2) with adjoint fermions,''
  Phys.\ Rev.\  D {\bf 80}, 074507 (2009)
  [arXiv:0907.3896 [hep-lat]].
  %%CITATION = PHRVA,D80,074507;%%

\bibitem{Bursa:2009we}
  F.~Bursa, L.~Del Debbio, L.~Keegan, C.~Pica and T.~Pickup,
  ``Mass anomalous dimension in SU(2) with two adjoint fermions,''
  Phys.\ Rev.\  D {\bf 81}, 014505 (2010)
  [arXiv:0910.4535 [hep-ph]].
  %%CITATION = PHRVA,D81,014505;%%

%\cite{Bursa:2009tj}
\bibitem{Bursa:2009tj}
  F.~Bursa, L.~Del Debbio, L.~Keegan, C.~Pica and T.~Pickup,
  ``Running of the coupling and quark mass in SU(2) with two adjoint
  fermions,''
  arXiv:0910.2562 [hep-ph].
  %%CITATION = ARXIV:0910.2562;%%

 %\cite{Hietanen:2009zz}
\bibitem{Hietanen:2009zz}
  A.~Hietanen, J.~Rantaharju, K.~Rummukainen and K.~Tuominen,
  ``Minimal technicolor on the lattice,''
  Nucl.\ Phys.\  A {\bf 820}, 191C (2009).
  %%CITATION = NUPHA,A820,191C;%%

%\cite{Hietanen:2009az}
\bibitem{Hietanen:2009az}
  A.~J.~Hietanen, K.~Rummukainen and K.~Tuominen,
  ``Evolution of the coupling constant in SU(2) lattice gauge theory with two
  adjoint fermions,''
  Phys.\ Rev.\  D {\bf 80}, 094504 (2009)
  [arXiv:0904.0864 [hep-lat]].
  %%CITATION = PHRVA,D80,094504;%%
 
%\cite{Catterall:2009sb}
\bibitem{Catterall:2009sb}
  S.~Catterall, J.~Giedt, F.~Sannino and J.~Schneible,
  ``Probes of nearly conformal behavior in lattice simulations of minimal
  walking technicolor,''
  arXiv:0910.4387 [hep-lat].
  %%CITATION = ARXIV:0910.4387;%%
    
 
%\cite{Kiskis:2003rd}
\bibitem{Kiskis:2003rd}
  J.~Kiskis, R.~Narayanan and H.~Neuberger,
   ``Does the crossover from perturbative to non-perturbative physics in QCD
  become a phase transition at infinite N?,''
  Phys.\ Lett.\  B {\bf 574}, 65 (2003)
  [arXiv:hep-lat/0308033].
  %%CITATION = PHLTA,B574,65;%%   
    
    
%\cite{Unsal:2008ch}
\bibitem{Unsal:2008ch}
  M.~\"Unsal and L.~G.~Yaffe,
  ``Center-stabilized Yang-Mills theory: confinement and large $N$ volume independence,''
  Phys.\ Rev.\  D {\bf 78}, 065035 (2008)
  [arXiv:0803.0344 [hep-th]].
  %%CITATION = PHRVA,D78,065035;%%
  
  
  

%\cite{Bringoltz:2009kb}
\bibitem{Bringoltz:2009kb}
  B.~Bringoltz and S.~R.~Sharpe,
  ``Non-perturbative volume-reduction of large-N QCD with adjoint fermions,''
  Phys.\ Rev.\  D {\bf 80}, 065031 (2009)
  [arXiv:0906.3538 [hep-lat]].
  %%CITATION = PHRVA,D80,065031;%%


%\cite{Bietenholz:2006cz}
\bibitem{Bietenholz:2006cz}
  W.~Bietenholz, J.~Nishimura, Y.~Susaki and J.~Volkholz,
   ``A non-perturbative study of 4d U(1) non-commutative gauge theory: The fate
  of one-loop instability,''
  JHEP {\bf 0610}, 042 (2006)
  [arXiv:hep-th/0608072].
  %%CITATION = JHEPA,0610,042;%%

%\cite{Teper:2006sp}
\bibitem{Teper:2006sp}
  M.~Teper and H.~Vairinhos,
  ``Symmetry breaking In twisted Eguchi-Kawai models,''
  Phys.\ Lett.\  B {\bf 652}, 359 (2007)
  [arXiv:hep-th/0612097].
  %%CITATION = PHLTA,B652,359;%%

%\cite{Azeyanagi:2007su}
\bibitem{Azeyanagi:2007su}
  T.~Azeyanagi, M.~Hanada, T.~Hirata and T.~Ishikawa,
  ``Phase structure of twisted Eguchi-Kawai model,''
  JHEP {\bf 0801}, 025 (2008)
  [arXiv:0711.1925 [hep-lat]].
  %%CITATION = JHEPA,0801,025;%%

%\cite{Bringoltz:2008av}
\bibitem{Bringoltz:2008av}
  B.~Bringoltz and S.~R.~Sharpe,
  ``Breakdown of large-N quenched reduction in SU(N) lattice gauge theories,''
  Phys.\ Rev.\  D {\bf 78}, 034507 (2008)
  [arXiv:0805.2146 [hep-lat]].
  %%CITATION = PHRVA,D78,034507;%%


  
%\cite{Cossu:2009sq}
\bibitem{Cossu:2009sq}
  G.~Cossu and M.~D'Elia,
  ``Finite size phase transitions in QCD with adjoint fermions,''
  JHEP {\bf 0907}, 048 (2009)
  [arXiv:0904.1353 [hep-lat]].
  %%CITATION = JHEPA,0907,048;%%

  
%\cite{Bedaque:2009md}
\bibitem{Bedaque:2009md}
  P.~F.~Bedaque, M.~I.~Buchoff, A.~Cherman and R.~P.~Springer,
  ``Can fermions save large N dimensional reduction?,''
  JHEP {\bf 0910}, 070 (2009)
  [arXiv:0904.0277 [hep-th]].
  %%CITATION = JHEPA,0910,070;%%

 

%\cite{Bringoltz:2009mi}
\bibitem{Bringoltz:2009mi}
  B.~Bringoltz,
   ``Large-N volume reduction of lattice QCD with adjoint Wilson fermions at
  weak-coupling,''
  JHEP {\bf 0906}, 091 (2009)
  [arXiv:0905.2406 [hep-lat]].
  %%CITATION = JHEPA,0906,091;%%


%\cite{Bringoltz:2009fj}
\bibitem{Bringoltz:2009fj}
  B.~Bringoltz,
   ``Partial breakdown of center symmetry in large-N QCD with adjoint Wilson
  fermions,''
  JHEP {\bf 1001}, 069 (2010)
  [arXiv:0911.0352 [hep-lat]].
  %%CITATION = JHEPA,1001,069;%%

%\cite{Poppitz:2009fm}
\bibitem{Poppitz:2009fm}
  E.~Poppitz and M.~\"Unsal,
  ``Comments on large-N volume independence,''
  JHEP {\bf 1001}, 098 (2010)
  [arXiv:0911.0358 [hep-th]].
  %%CITATION = JHEPA,1001,098;%%



%\cite{Myers:2008zm}
\bibitem{Myers:2008zm}
  J.~C.~Myers and M.~C.~Ogilvie,
  ``Exotic phases of finite temperature SU(N) gauge theories,''
  Nucl.\ Phys.\  A {\bf 820}, 187C (2009)
  [arXiv:0810.2266 [hep-th]].
  %%CITATION = NUPHA,A820,187C;%%
  
%\cite{Meisinger:2009ne}
\bibitem{Meisinger:2009ne}
  P.~N.~Meisinger and M.~C.~Ogilvie,
  ``String Tension Scaling in High-Temperature Confined SU(N) Gauge Theories,''
  Phys.\ Rev.\  D {\bf 81}, 025012 (2010)
  [arXiv:0905.3577 [hep-lat]].
  %%CITATION = PHRVA,D81,025012;%%



%\cite{Hietanen:2009ex}
\bibitem{Hietanen:2009ex}
  A.~Hietanen and R.~Narayanan,
  ``The large N limit of four dimensional Yang-Mills field coupled to adjoint
  fermions on a single site lattice,''
  arXiv:0911.2449 [hep-lat].
  %%CITATION = ARXIV:0911.2449;%%

%\cite{Azeyanagi:2010ne}
\bibitem{Azeyanagi:2010ne}
  T.~Azeyanagi, M.~Hanada, M.~\"Unsal and R.~Yacoby,
  ``Large-N reduction in QCD-like theories with massive adjoint fermions,''
  arXiv:1006.0717 [hep-th].
  %%CITATION = ARXIV:1006.0717;%%
  
\bibitem{hmc} S. Duane, A. Kennedy, B. Pendleton and D. Roweth, Phys. Lett.
B195B (1987) 216. 




\end{thebibliography}

\end{document}